\documentclass{webofc}

\usepackage[varg]{txfonts}   
\usepackage{hyperref}
\usepackage{url}
\usepackage{graphicx}
\usepackage[font=small,labelfont=bf,labelsep=colon]{caption}
\hypersetup{colorlinks=true,citecolor=blue,urlcolor=blue,linkcolor=blue}
\begin{document}
\title{Neutrino Program at Fermilab - Enhancing proton beam power and accelerator infrastructure}
%
%

\author{\firstname{Ganguly} \lastname{Sudeshna}\thanks{\email{sganguly@fnal.gov}} \\
        \textit{On behalf of the Accelerator Directorate at Fermilab}
        }

\institute{Fermilab,  Batavia, 60510, IL, US
          }

\abstract{The upcoming long baseline neutrino experiments aim to enhance proton beam power to multi-MW scale and utilize large-scale detectors to address the challenge of limited event statistics. The DUNE experiment at LBNF will test the three neutrino flavor paradigm and directly search for CP violation by studying oscillation signatures in the high intensity $\nu_{\mu}$ (anti-$\nu_{\mu}$) beam to $\nu_{e}$ (anti-$\nu_{e}$) measured over a long baseline.
Higher beam power and improved accelerator up-time will enhance neutrino flux for the neutrino program by increasing the number of protons on target. LBNF/DUNE, as well as PIP-II upgrade and Accelerator Complex Evolution (ACE) plan, play a vital role in this effort. The scientific potential of ACE plan extends beyond neutrino physics, encompassing endeavors such as the Muon Collider, Charged Lepton Flavor Violation (CLFV), Dark Sectors, and exploration of neutrinos beyond DUNE.\par
In the era of higher-power accelerator operation, research in target materials and beam instrumentation is crucial for optimizing design modifications.
This abstract discusses Fermilab ACE, the science opportunities it provides, and how Fermilab is pushing the limits of proton beam power and accelerator infrastructure. By tackling neutrino beam challenges and exploring research and development ideas, we are advancing our understanding of fundamental particles and their interactions.

}
\maketitle
\section{Introduction}
\label{intro}
\begin{figure}[h]
\centering
\includegraphics[width=9cm,clip]{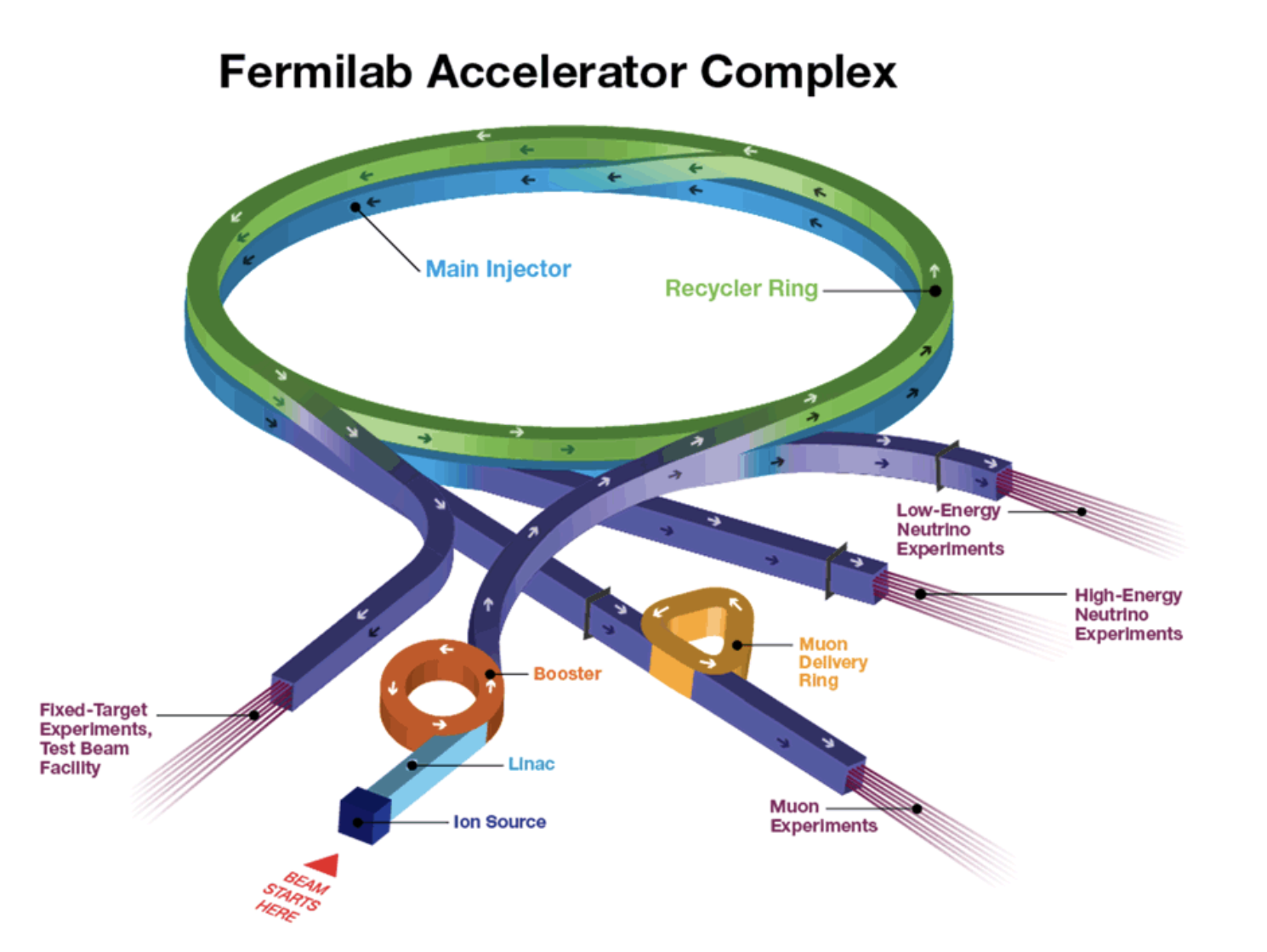}
\captionsetup{justification=centering}
    \caption{Present Fermilab Accelerator complex.}
\label{fig-1}       
\end{figure}

Key components of the current Fermilab accelerator complex include the Linear Accelerator (Linac), the Booster, the Main Injector (MI), and the Recycler ring (RR). The Main Injector is central to the complex, capable of accelerating protons to high energies before they are directed to various experiments. The present Fermilab complex is shown in figure \ref{fig-1}.
An ion source generates H- ions that are accelerated to 400 MeV by the Linac. Upon injection into the Booster, the H- ions are converted to protons. The present Booster is a rapid cycling synchrotron with a repetition rate of 15 Hz. The Booster accelerates protons to 8 GeV and delivers them into the Recycler ring for high energy neutrino experiments and the muon campus or to the Booster Neutrino Beam experiments. The Booster delivers 12 batches of beam into the Recycler which doubles the beam intensity by a RF beam manipulation known as slip-stacking to combine bunches effectively at 8 GeV and sends the beam to the Main Injector which accelerates the beam to 120 GeV and delivers it to the Neutrinos at the Main Injector (NuMI) beamline. The Proton Improvement Plan II (PIP-II) will provide new superconducting radio-frequency (SRF) Linacs for injection into the Booster at 800 MeV. PIP-II will increase the booster cycle rate from 15 Hz to 20 Hz. MI will increase proton beam intensity for 1.2 MW of beam power for Deep Underground Neutrino Experiment (DUNE) at the Long Baseline Neutrino Facility (LBNF).\par
In 2023, Fermilab hosted an international workshop on Accelerator Complex Evolution (ACE) physics \cite{Gori2024}. There are two main phases to ACE:
\begin{itemize}
    \item ACE-MIRT (Main Injector Ramp and Target) upgrade: In this phase, after PIP-II is online, the MI ramp time for DUNE will be shortened and the beam power will be increased to 2.1 MW. R\&D is currently being conducted on the target to ensure it can handle up to 2.4 MW.
    \item ACE-BR (Booster Replacement) upgrade: In this later phase, the Booster synchrotron will be replaced to provide 2.4 MW of beam power to DUNE, which will facilitate next-generation accelerator experiments.
\end{itemize}
The ACE-MIRT plan strives to accelerate the delivery of DUNE science. As an example, we can achieve mass ordering sensitivity of 5$\sigma$ after 3.5 years for 100\% of $\delta_{CP}$ values instead of 5 years.\par
DUNE physics goals are dependent on High Power Targetry during the ACE-MIRT era. Beam instrumentation is also crucial for the smooth operation of an accelerator complex, as it integrates various diagnostic and beam control systems. Currently, four pixelated ionization chamber detectors \cite{Yonehara2023} are located downstream of NuMI beamline decay pipe. These detectors are employed to monitor the health of the target by observing the profiles of secondary and tertiary particles. Unfortunately, the first ion chamber, positioned closest to the target, has endured considerable radiation damage, rendering it unable to fulfill its intended function of reconstructing proton beam intensity and predicting target deterioration. The development of radiation-resistant beam detectors could mitigate these radiation-induced problems.

\section{NuMI Megawatt upgrade project}
\label{sec-1}
NuMI Megawatt Upgrade at Fermilab aimed at enhancing beam power involved a series of significant upgrades. Through the NOvA/Accelerator and NuMI Upgrades (ANU), the beam power was increased to 700 kW. Between 2018 and 2021, the NuMI Megawatt Accelerator Improvement Project (AIP) upgraded the system to handle nearly 1 MW beam power. Target, horns, and supporting systems were upgraded, requiring three annual shutdowns for component replacements. An important milestone was achieved in May 2023 when a record beam power of nearly 959 kW was achieved with a main injector cycle of only 1.133 seconds. To push towards a 1 MW beam, the Main Injector was upgraded to operate at 1.067 s in 2024. In NuMI, MI cycle times were reduced and components on the target station were upgraded so that the beam power could be expanded to nearly 1 MW with the existing machines.

\section{Accelerator Complex Evolution (ACE)}
\label{sec-2}
\begin{figure}[h]
\centering
\includegraphics[width=9cm,clip]{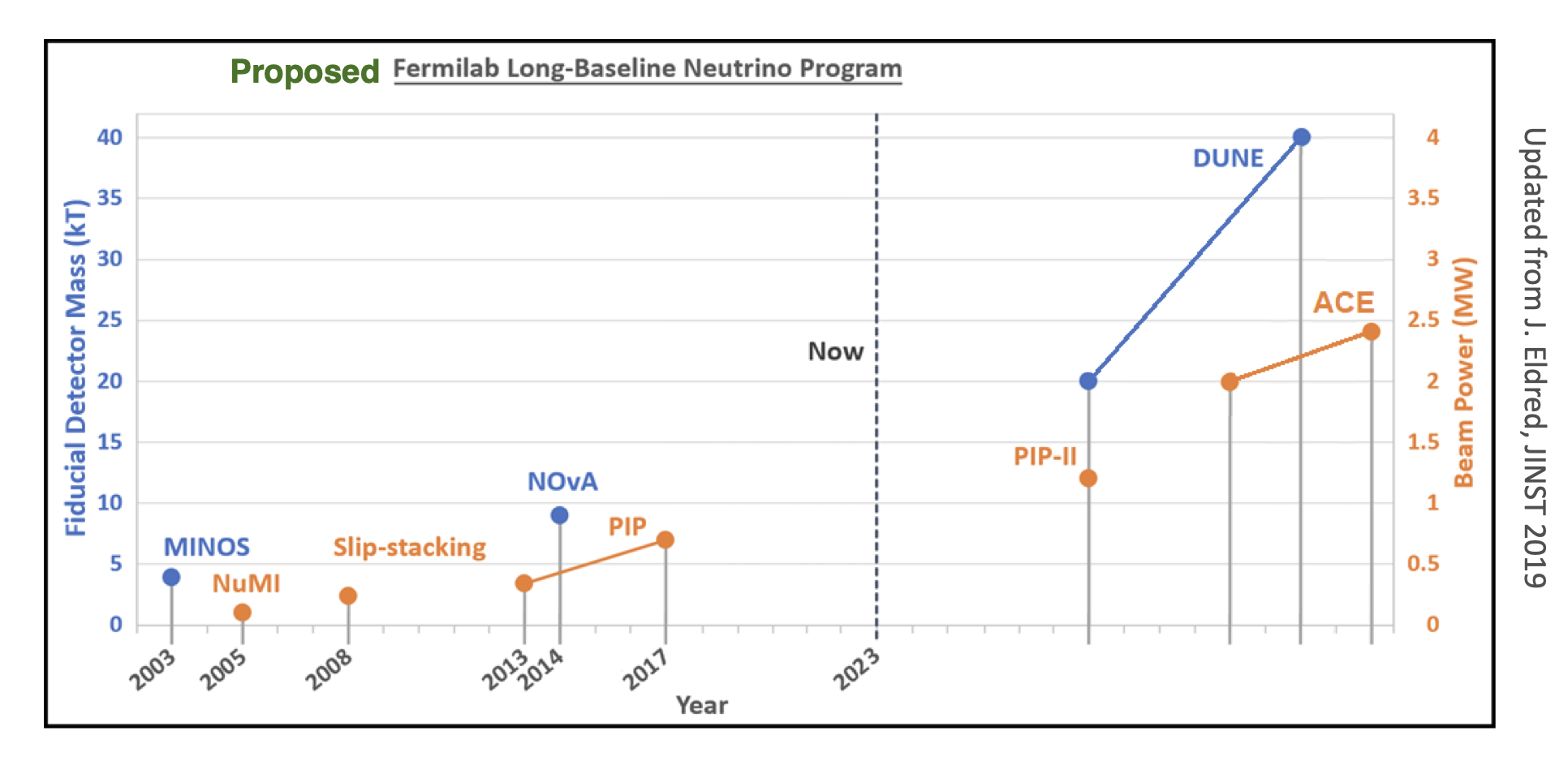}
\captionsetup{justification=centering}
    \caption{Long Baseline Neutrino Programs (with ACE).}
\label{fig-2}       
\end{figure}
The ACE-MIRT project aims to increase beam power to over 2 MW by optimizing the Main Injector cycle time. LBNF and DUNE are Fermilab's top priorities, which can be achieved by maximizing neutrino production. In the ACE-MIRT initiative, the Main Injector ramp time will be reduced to approximately 0.65 seconds (the absolute minimum slip stacking time is 0.65 seconds), increasing beam power. It complements the Proton Improvement Plan-II (PIP-II), which targets 1.2 MW with a 1.2-second cycle.
Figure \ref{fig-2} \cite{Eldred2023} shows the progression from current capabilities, with PIP-II upgrades targeting a beam power of 1.2 MW, to the proposed ACE upgrades that aim to exceed 2 MW by reducing the Main Injector (MI) cycle time. The equation 
\[ P = \frac{eNE}{T} \]
illustrates the relationship between beam power (\(P\)), the number of particles (\(N\)), the energy per particle (\(E\)), and the cycle time (\(T\)). By decreasing \(T\), ACE-MIRT aims to significantly increase beam power.
A 0.65 second cycle, however, poses challenges, such as the need for additional infrastructure, and it leaves no beam for the 8-GeV program. 
\begin{figure}[h]
\centering
\includegraphics[width=13cm,clip]{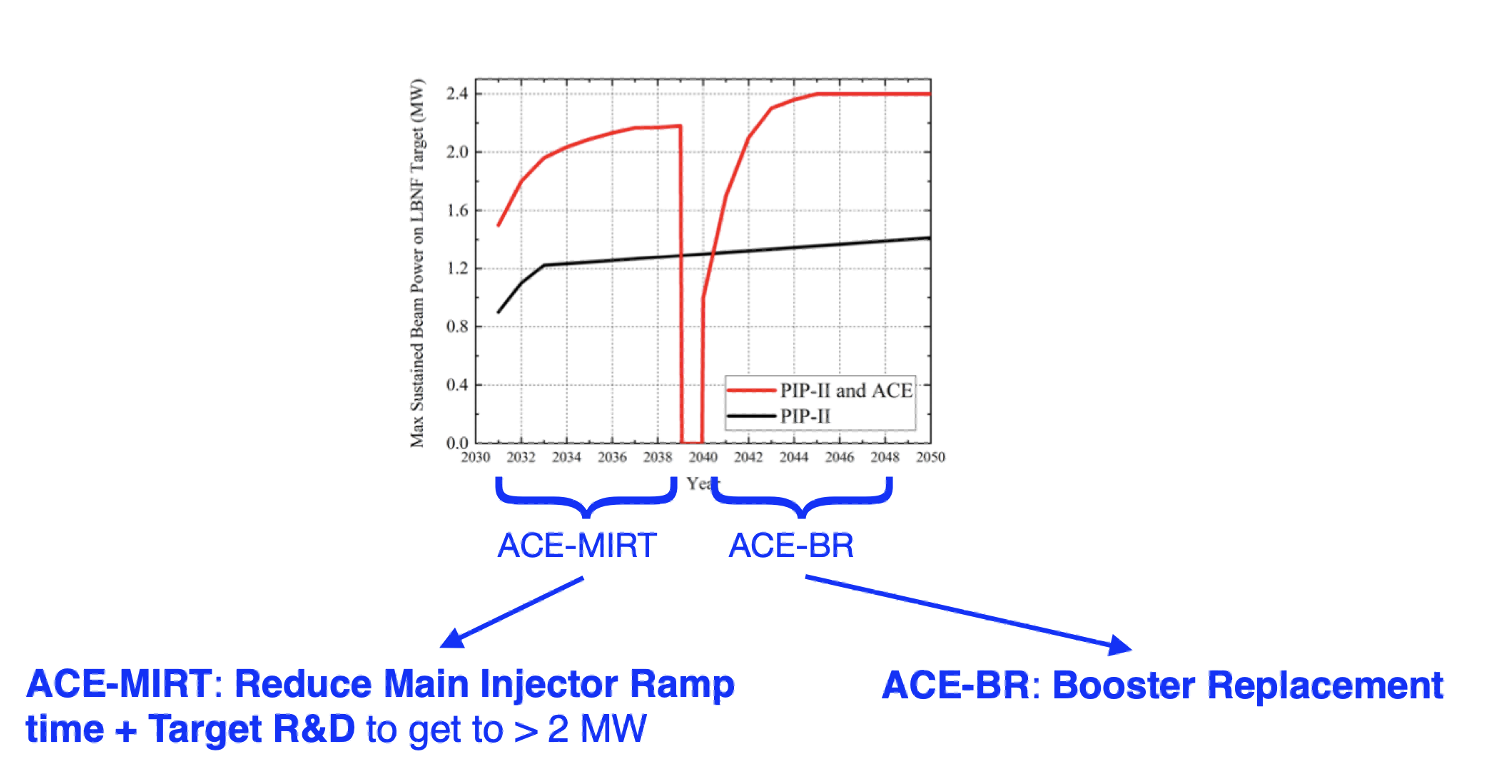}
\captionsetup{justification=centering}
    \caption{Dune Plan (with ACE).}
\label{fig-3}       
\end{figure}
Figure \ref{fig-3} \cite{Fermilab2023} illustrates the projected sustained beam power on the LBNF target from 2030 to 2050, comparing the effects of PIP-II alone versus the combined PIP-II and ACE initiatives. The black line represents PIP-II alone, showing a steady beam power of approximately 1.2 MW. In contrast, the red line, which includes the ACE initiatives, shows an increase in power reaching around 2 MW by 2038, followed by a significant rise to over 2 MW by 2042. The ACE-MIRT phase focuses on reducing Main Injector ramp time and enhancing target R\&D, while the ACE-BR phase involves booster replacement, both crucial for achieving higher beam power.
\begin{figure}[h]
\centering
\includegraphics[width=9cm,clip]{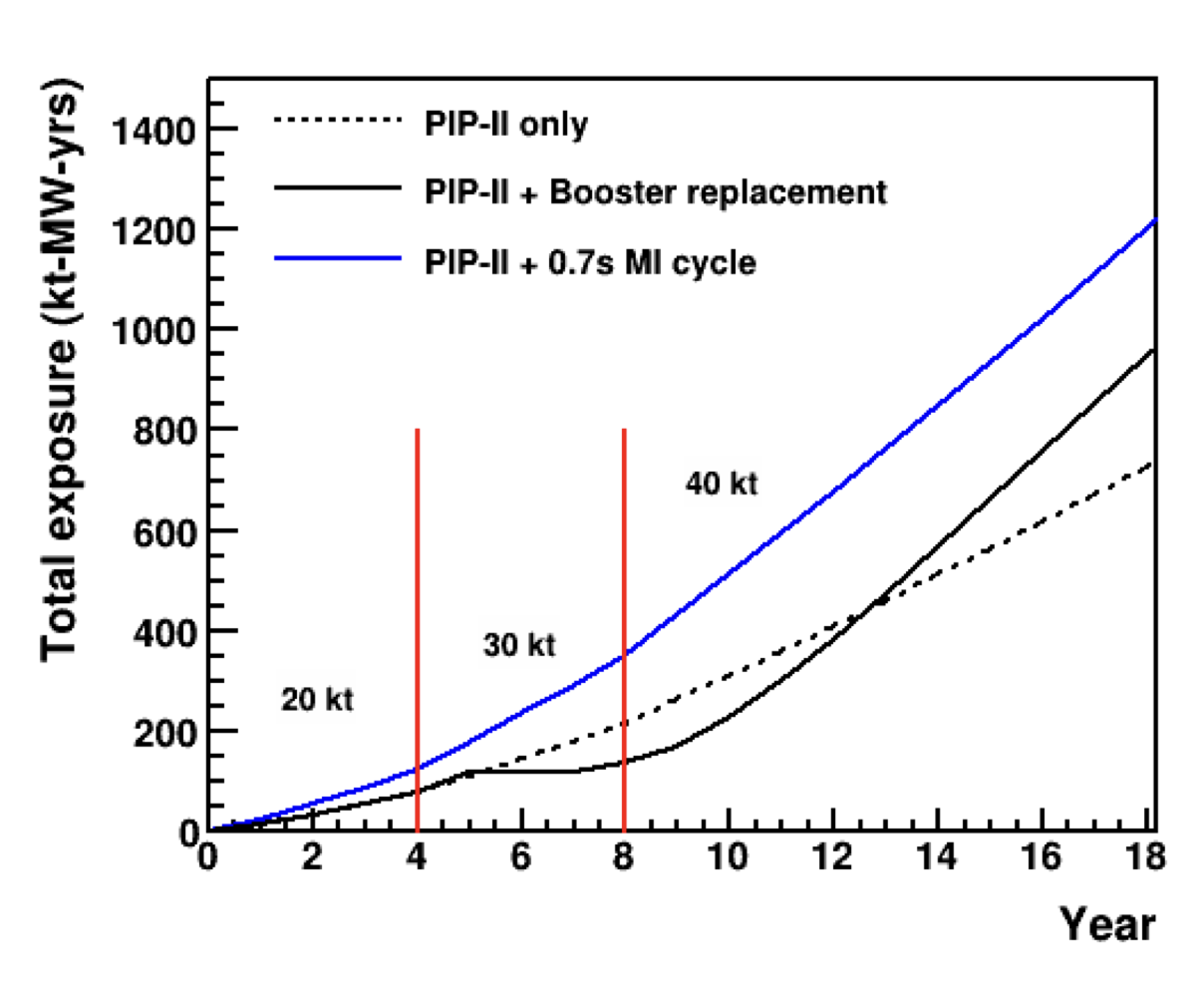}
\captionsetup{justification=centering}
    \caption{Projected Total Exposure for DUNE with Different Accelerator Upgrades.}
\label{fig-4}       
\end{figure}
Figure \ref{fig-4} \cite{Marshall2023} illustrates the projected impact of ACE on DUNE sensitivities, emphasizing the relationship between exposure (measured in kiloton-megawatt-years) and years. The figure highlights the significant role of ACE in optimizing the 40 kT DUNE detector by increasing the beam power to over 2 MW. ACE-MIRT option or "PIP-II + 0.7s MI Cycle" scenario (solid blue line) provides the fastest and most significant increase in total exposure, reducing the Main Injector cycle time to 0.65 seconds.


%
%
%
\section{Neutrino Beam Challenges}
\label{sec-4}
Beam power at major neutrino beam facilities is often limited by target survivability rather than accelerator capability. For facilities to operate at higher beam powers, high-power targetry research and development (R\&D) is essential. Radiation-induced swelling occurred on the MINOS NT-02 target and fractures occurred on the NOvA MET-01 target \cite{Frederique2023}, highlighting the challenges of material degradation. As part of the NuMI megawatt upgrade project, the target core had to be redesigned for an increased beam spot size from 1.3 mm to 1.5 mm, with winged fins added upstream to account for beam miss-steering and a cooling loop added downstream. It is imperative to evaluate current data to accurately predict target lifetimes when accelerating upgrades to 2+MW. To ensure the sustainability of high-power operations, new facilities for irradiation stations, Post-Irradiation Examination (PIE), and advanced modeling are essential.\par
In accelerator complexes, radiation-hardened beam instrumentation is crucial for ensuring reliable and smooth operation, as it reduces the adverse effects of radiation exposure, temperature, and humidity. Especially in high-radiation environments, these robust instruments are crucial to maintaining accurate measurements and controlling beam parameters, preventing damage and guaranteeing consistency. By improving beam monitors, we'll be able to run higher beam power, which is important for future multi-megawatt facilities. 
\section{Novel instrumentation ideas}
A high-intensity muon neutrino beam is created at Fermilab's NuMI facility by colliding protons with carbon targets. Two focusing horns direct 120-GeV proton beams into a cylindrical decay pipe. As a result of this process, mesons can be converted to neutrinos and muons through decay. A series of four-layered pixelated ionization chambers are situated downstream of a decay pipe. By analyzing secondary and tertiary particle profiles, these chambers monitor the overall health of the target system. The three chambers known as the "muon monitor (MM)" are strategically shielded from intense radiation by a thick material that effectively absorbs the majority of charged particles, save for muons. A high-power proton beam, however, causes a significant increase in muon flux at the monitor, decreasing its performance due to a space charge effect. 
As a solution to this challenge, a Large Area Picosecond Photodetector (LAPPD)-based \cite{Angelico2019} radiation-hardened muon monitor can be placed in the muon alcoves following the absorber in the NuMI beamline, allowing us to precisely measure the muon TOF across the transverse plane using a bunch-by-bunch proton beam approach. For example, in the NuMI beamline, the highest recorded muon fluence is approximately $10^6$ per square centimeter per spill at the muon alcove 2 location. By repositioning the LAPPD away from the beam center, the fluence could be significantly reduced by about two orders of magnitude, allowing the LAPPDs to withstand such radiation levels.
Existing muon monitors in the NuMI facility consist of a $9x9$ array of ionization chambers measuring the integrated muon flux, but it lacks timing information. LAPPDs can serve as an effective muon monitor not only in the NuMI beamline but also in upcoming high-power neutrino oscillation experiments.\par
\begin{figure}[h]
\centering
\includegraphics[width=9cm,clip]{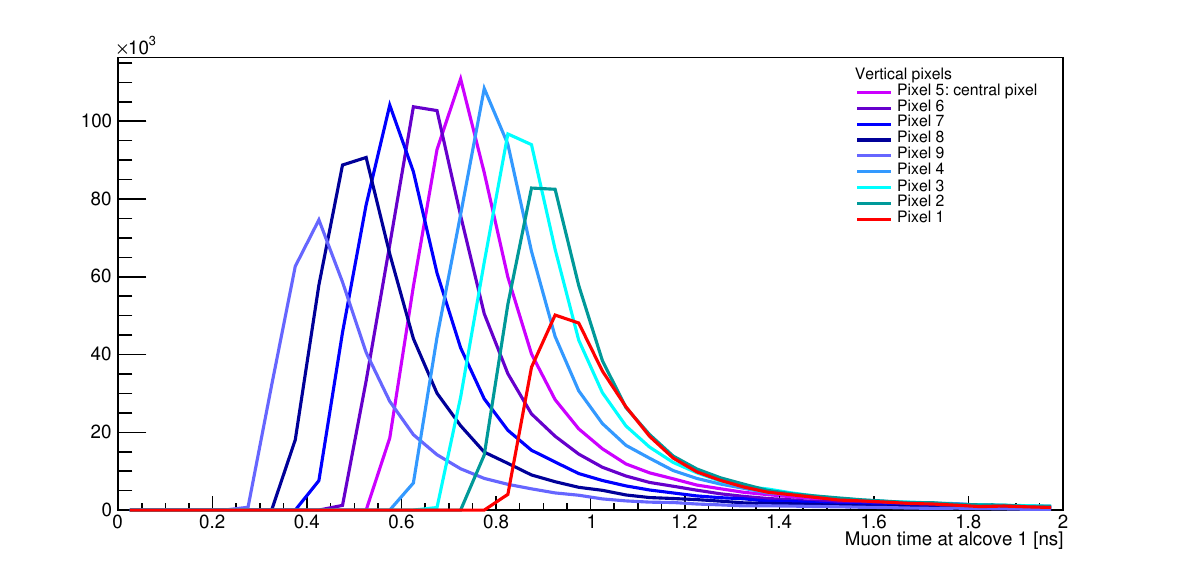}
\captionsetup{justification=centering}
    \caption{Muon time distributions at MM1 for different vertical pixels.}
\label{fig-5}       
\end{figure}
With simulated data, Figure \ref{fig-5} illustrates how individual pixel of MM1 sees a different muon spectrum. This characteristic of the muon monitor data not only works as a target health monitor but is pivotal for analyzing the impact of chromaticity on magnetic horn focusing and bench-marking beamline data against simulation \cite{Yonehara2023}. 
\begin{figure}[h]
\centering
\includegraphics[width=9cm,clip]{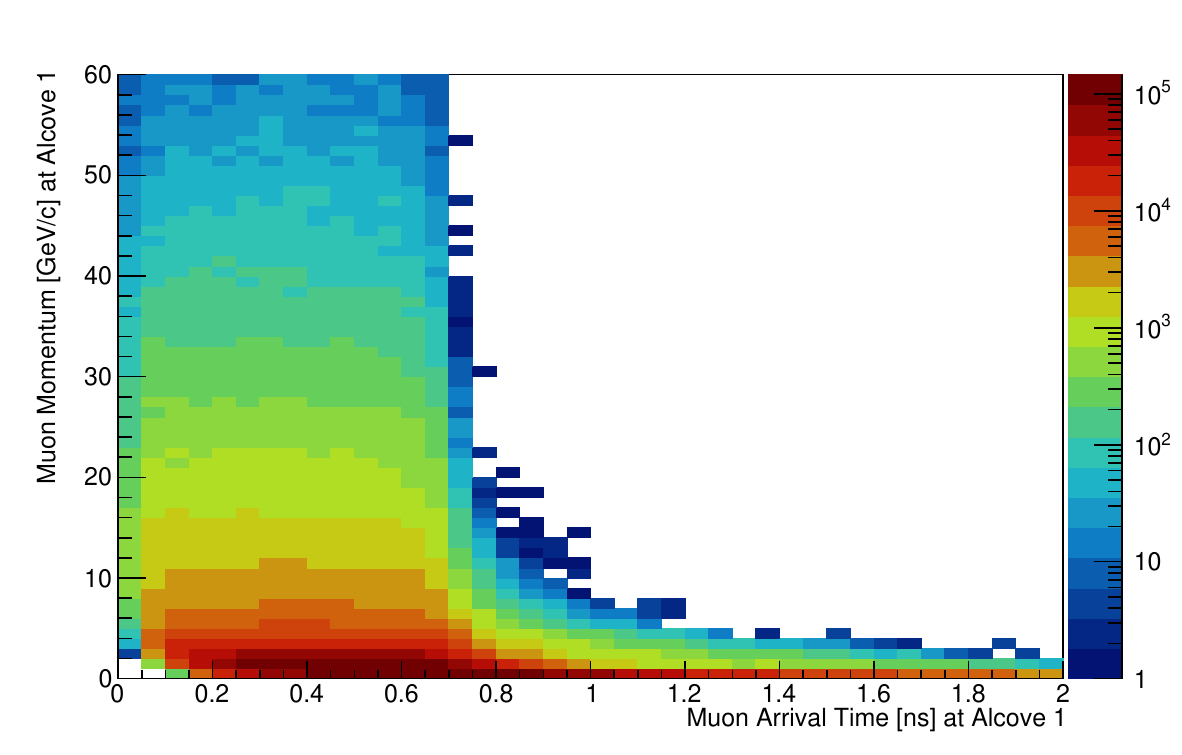}
\captionsetup{justification=centering}
    \caption{Simulated time-of-flight vs muon momentum at MM1.}
\label{fig-6}       
\end{figure}
At MM1, the simulated time-of-flight (TOF) of muons is plotted against muon momentum in figure \ref{fig-6}. The use of LAPPDs offers a practical solution for measuring muon TOF in muon alcoves. 

\begin{figure}[h]
\centering
\includegraphics[width=9cm,clip]{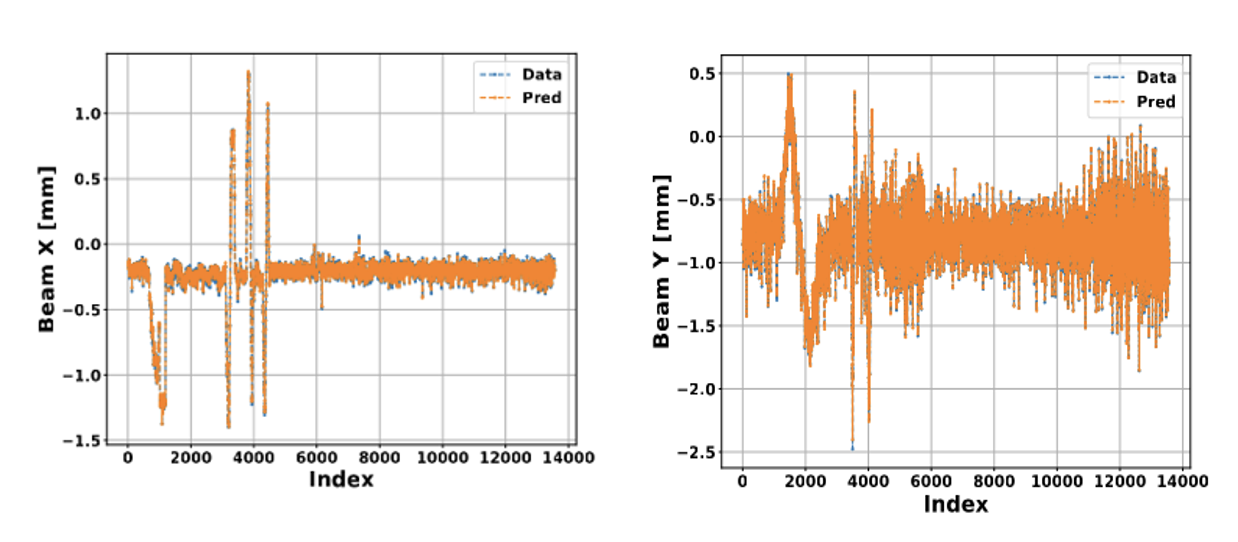}
\captionsetup{justification=centering}
    \caption{Beam position predictions using ANN model with muon monitor data as input. }
\label{fig-7}       
\end{figure}

LAPPDs already offer a space resolution of 1x1 mm and a time resolution of ~55 ps or better, making them ideal for this purpose.\par
Machine  learning (ML) techniques \cite{Wickremasinghe2023} have been employed at Fermilab to assess the pixelated images generated by the muon monitors, enabling precise reconstruction of the proton beam's position at the target and the horn current with a measurement uncertainty of less than 1\%. The NuMI horn's linear beam optics nature suggests a linear relationship between the muon monitor response and beam parameter variations. 

Initial studies on the muon monitor data from the NuMI beamline have shown that ML algorithm, using an Artificial Neural Network with multiple hidden layers, successfully predicts beam positions at the target based on 241 observed values, with an accuracy of ±0.018 mm horizontally and ±0.013 mm vertically. These results demonstrate that ML can reliably detect parameters crucial for the NOvA experiment, on par with traditional instrumentation accuracy. ML also has potential applications in identifying anomalies, but its accuracy relies on quality training data and system optimization. Figure \ref{fig-7} shows an example of predicting the beam position on the NuMI target, where the input comprises signals from the muon monitor pixels. 

\section{Summary}
\label{sec-5}
The ACE-MIRT initiative aims to upgrade the Main Injector to reduce ramp time and deliver increased beam power to the DUNE experiment, targeting a maximum of approximately 2.1 MW as quickly as possible. This effort necessitates extensive R\&D on target technologies to ensure DUNE can manage up to 2.4 MW of beam power. 
Additionally, significant research is required to develop radiation-hardened beam instrumentation to support these high-power operations. 

\section*{Acknowledgements}
This work is supported by the Fermi Research Alliance, LLC manages and operates the Fermi National Accelerator Laboratory pursuant to Contract number DE-AC02-07CH11359 with the United States Department of Energy.

This work is partially supported by the U.S. Department of Energy grant DE-SC0019264.

\end{document}